\documentclass[aps,prl,twocolumn,superscriptaddress,groupedaddress,tightenlines]{revtex4}

\usepackage{amssymb}
\usepackage{color,graphicx}
\usepackage{amsmath}
\usepackage{amsbsy}
\usepackage{amsthm}
\usepackage{bbm}
\usepackage{bm}
\usepackage{epsfig}
\usepackage{lscape}
\usepackage{float}
\usepackage{subfigure}

\begin{document}

\title{Surpassing the Energy Resolution Limit with ferromagnetic torque sensors}

\author{Andrea Vinante}
\email{anvinante@fbk.eu}
\affiliation{Istituto di Fotonica e Nanotecnologie – CNR and Fondazione Bruno Kessler, I-38123 Povo, Trento, Italy}

\author{Chris Timberlake}
\affiliation{Department of Physics and Astronomy, University of Southampton, Southampton
SO17 1BJ, UK}

\author{Dmitry Budker}
\affiliation{Johannes Gutenberg-Universität Mainz, 55128 Mainz, Germany}
\affiliation{Helmholtz-Institute, GSI Helmholtzzentrum für Schwerionenforschung, 55128 Mainz, Germany}
\affiliation{Department of Physics, University of California at Berkeley, Berkeley, CA 94720-7300, US}

\author{Derek F. Jackson Kimball}
\affiliation{Department of Physics, California State University-East Bay, Hayward, CA 94542-3084, US}

\author{Alexander O. Sushkov}
\affiliation{Department of Physics, Boston University, Boston, MA 02215, US}
\affiliation{Department of Electrical and Computer Engineering, Boston University, Boston, MA 02215, USA}
\affiliation{Photonics Center, Boston University, Boston, MA 02215, USA}

\author{Hendrik Ulbricht}
\affiliation{Department of Physics and Astronomy, University of Southampton, Southampton
SO17 1BJ, UK}

\date{\today}

\begin{abstract}
We discuss the fundamental noise limitations of a ferromagnetic torque sensor based on a levitated magnet in the tipping regime. We evaluate the optimal magnetic field resolution taking into account the thermomechanical noise and the mechanical detection noise at the standard quantum limit (SQL). We find that the Energy Resolution Limit (ERL), pointed out in recent literature as a relevant benchmark for most classes of magnetometers, can be surpassed by many orders of magnitude. Moreover, similarly to the case of a ferromagnetic gyroscope, it is also possible to surpass the standard quantum limit for magnetometry with independent spins, arising from spin-projection noise. Our finding indicates that magnetomechanical systems optimized for magnetometry can achieve a magnetic field resolution per unit volume several orders of magnitude better than any conventional magnetometer. We discuss possible implications, focusing on fundamental physics problems such as the search for exotic interactions beyond the standard model.
\end{abstract}

\maketitle

The precision of measurement devices in different contexts is ultimately constrained by quantum mechanics. For spin-based magnetic sensing, the uncertainty set by the Heisenberg principle on the magnetic field measured over a time $t$ by an ensemble of $N$ independent particles with gyromagnetic ratio $\gamma_0$ is \cite{kimball1}:
\begin{equation}
 \delta B \approx \frac{1}{\gamma_0} \sqrt {\frac{\Gamma_{\textrm{rel}}}{N t}},   \label{SQLmag}
\end{equation}
with the spin relaxation rate $\Gamma_{\textrm{rel}}$ to be replaced by $1/t$ in the limit $\Gamma_{\textrm{rel}} \rightarrow 0$. Equation~(\ref{SQLmag}) is known as spin-projection noise, and results in the so-called standard quantum limit for independent particles.

A different approach to assessing quantum limitations on magnetic field sensing has been recently discussed  \cite{mitchell} in terms of an energy resolution limit (ERL). The ERL relates the magnetic field resolution, the measurement time and the spatial resolution. It can be expressed by the simple formula:
\begin{equation}
  \frac{S_B}{2 \mu_0 } V \geq  \hbar  \label{ERL} ,
\end{equation}
where $S_B \left( \omega \right)$ is the power spectral density of the magnetic field noise at angular frequency $\omega$, $V$ is the sensing volume, $\hbar$ is the Planck's constant, $\mu_0$ is the vacuum permeability. The ERL is supported by several technology-specific arguments although, in contrast with Eq.~(\ref{SQLmag}), there is no rigorous argument supporting its general validity. Nevertheless, it has been empirically confirmed across a large number of technological platforms and over many orders of magnitudes of sensing volume. In particular, the most sensitive existing magnetometers, including SQUIDs, optically pumped atomic magnetometers, and solid state spins, can approach the ERL in the best cases \cite{mitchell}. In this respect, the ERL can be considered to be a useful and physically motivated benchmark to compare different classes of magnetometers.

However, the ERL should not be regarded as a fundamental limit, and several systems have been indeed suggested to surpass it \cite{mitchell}. In this Letter we show that there is a remarkable system offered us by nature, which can surpass the ERL by many orders of magnitude: a hard ferromagnet, in which a large number of spins are effectively locked together by magnetic anisotropy. 

A hard ferromagnet can be employed as a magnetic field or field-gradient sensor by measuring its mechanical response to an applied magnetic field. This is typically done by attaching the magnet to a mechanical oscillator. Well-known applications are magnetic force microscopy \cite{MFM} and magnetic resonance force microscopy \cite{MRFM}. Even more effectively, one can detect magnetic fields by levitating the magnet \cite{kimball1}. For small enough size and slow enough rotational motion, the angular momentum of a levitated ferromagnet is dominated by its intrinsic spin, and the rotational dynamics is gyroscopic. In this regime the total spin undergoes Larmor precession in an external field, similarly to free atoms. The fact that a ferromagnetic gyroscope can beat standard quantum limits on magnetometry was pointed out in Ref. \cite{kimball1}. Following this suggestion, ferromagnetic gyroscopes have been recently proposed as exquisite ultrasensitive sensors for fundamental physics, for instance to probe the general relativistic drag on quantum spin angular momentum \cite{kimball3} or to detect exotic interactions mediated by axionlike particles \cite{kimball4}.

In this Letter we analyze a more common and easily achievable scenario of a levitated ferromagnet in the tipping regime, where gyroscopic effects are negligible and the dynamics is purely librational. This is usually the case for micromagnets and macroscopic magnets. We take into account as unavoidable fundamental noise sources on a torque measurement the mechanical thermal noise and the quantum detection noise. Assuming as a benchmark for the latter the Standard Quantum Limit on mechanical measurements \cite{braginsky}, we find that the ERL can be surpassed by many orders of magnitude.

\begin{figure}[!ht]
\includegraphics[width=8.6cm]{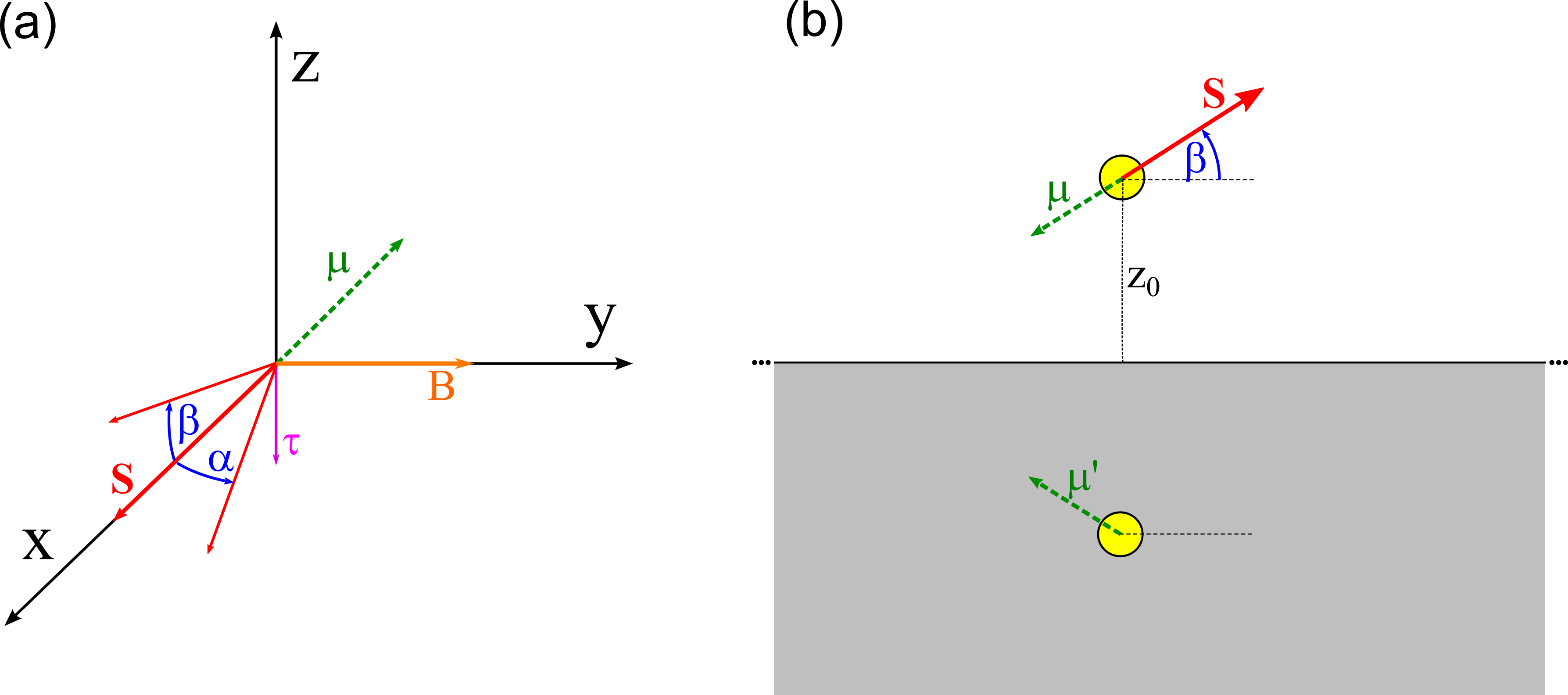}
\caption{(a) Model and conventions adopted. A ferromagnet with spin $\bm S$ and magnetic moment $\bm \mu=-\gamma_0 \bm S$ is trapped along the $x$ axis. It can harmonically librate around $z$ by an angle $\alpha$ and around $y$ by an angle $\beta$. Application of an oscillating magnetic field $\bm B$ along $y$ will produce a torque $\bm \tau=\bm \mu\times \bm B$ along $z$, driving the librational motion in the $\alpha$ degree of freedom. (b) Scheme of a ferromagnet above an infinite superconducting plane (gray region) in the Meissner regime. In this implementation, harmonic confinement of the $\beta$ angle arises naturally due to the interaction with an image dipole $\bm \mu'$.}
\end{figure}

Consider a levitated hard ferromagnet, for simplicity we take a sphere as in Refs.~\cite{gieseler,kimball2,maglev}. Let $\bm{\mu}=-\gamma_0 \bm{S}$ be the magnetic dipole moment, where $\bm{S}$ is the total spin angular momentum and $\gamma_0$ is the gyromagnetic ratio. We assume that, due to the nature of the levitating mechanism, the magnetic moment is trapped along the $x$ direction (Fig. 1a), with harmonic restoring torque for small librations around $z$ ($\alpha$ angle) and $y$ ($\beta$ angle) given by $\tau_z=-I \omega_\alpha^2 \alpha$ and $\tau_y=I \omega_\beta^2 \beta$. Here, $I$ is the isotropic moment of inertia and $\omega_\alpha$, $\omega_\beta$ are resonance frequencies of the librational modes. For small angles $\alpha$,$\beta$ in the linear approximation we can write the total spin as $\bm S \simeq  S \left( 1,\alpha,\beta \right)$ and the angular velocity vector as $\bm \Omega\simeq \left(\dot \gamma, -\dot \beta, \dot \alpha \right)$ where we have introduced the angle $\gamma$ corresponding to a rotation around $\bm S$. Finally, we assume that the magnet center of mass is in a stable equilibrium position and neglect cross-coupling between translational and librational degrees of freedom.

The situation described by this model is approximately the case of recent experimental implementations in which a magnetic microsphere is levitated either above a type II \cite{gieseler,kimball2} or a type I \cite{maglev} superconductor. In the former case translational and librational trappings are produced by pinning to vortices. In the latter, they are determined by the geometry of the trap. For instance, for a magnetic dipole placed above a superconducting plane in the absence of vortices, the horizontal position $\beta=0$ is found to be stable (Fig.~1b). The restoring torque is provided by the Meissner field, which is equivalent to that produced by an image dipole. Analytical calculations \cite{maglev} show that the Meissner field is given by:
\begin{equation}
  \bm {B_i} =-\frac{\mu_0}{32 \pi z_0^3}\left( \mu_x, \mu_y, 2 \mu_z \right),
\end{equation}
where $z_0$ is the equilibrium height above the superconductor.
For small $\beta$ the field magnitude is $B_i\simeq \mu_0 \mu/\left( 32 \pi z_0^3 \right)$ and the restoring frequency can be written as:
\begin{equation}
  \omega^2_\beta=\omega_{L}\omega_I ,
\end{equation}
where $\omega_{L} = \gamma_0 B_i$ is the Larmor frequency in the field $B_i$ and $\omega_I=S/I$ is the Einstein-de Haas frequency associated with the spin angular momentum $\bm S$. For the $\alpha$ degree of freedom, trapping may arise for different reasons. For instance, in Ref.~\cite{maglev} the $\alpha$ motion was found to be trapped despite the cylindrical geometry of the trap, owing to a small symmetry breaking induced by tilt effects.
 
Following Ref.~\cite{kimball3} we write the angular equations of motion of the dipole as:
\begin{align} 
\dot {\bm J} &= \bm {\tau_i}+ \bm {\tau} , \\
\dot{\bm S} &= \bm \Omega \times \bm S,   \label{eq5}
\end{align}
where $\bm J = \bm L + \bm S$ is the total angular momentum with $\bm L = I \bm \Omega$ being the kinetic angular momentum, $\bm \tau_i = \bm \mu \times \bm {B_i}$ is the restoring torque of the Meissner field $\bm {B_i}$, and $\bm \tau$ is an external torque. The latter could be induced by an additional field $\bm B$. Equation~(\ref{eq5}) strictly holds for hard ferromagnets, i.e., by assuming that the spin is always locked to the anisotropy axis and is therefore rigidly attached to the magnet crystal \cite{kimball3,romero}. 

In the linearized limit the equations of motion can be expressed in terms of the angles $\alpha$ and $\beta$ as:
\begin{align} 
\ddot \alpha &= - \omega_\alpha^2 \alpha - \frac{\omega_\alpha}{Q_\alpha}\dot \alpha-(\omega_I+\dot \gamma) \dot \beta +\frac{\tau_z}{I} ,   \label{eqalpha}\\
\ddot \beta &= - \omega_\beta^2 \beta -\frac{\omega_\beta}{Q_\beta}\dot\beta+(\omega_I+\dot \gamma) \dot \alpha -\frac{\tau_y}{I},       \label{eqbeta}  
\end{align}
which is a system of two kinetically coupled harmonic oscillators. Here, damping has been added by hand by defining phenomenological quality factors $Q_\alpha$ and $Q_\beta$. The cross-coupling kinetic terms are characteristic of gyroscopic behaviour. They contain a classical gyroscopic term proportional to $\dot \gamma$, and a Einstein-de Haas term proportional to $\omega_I$ that expresses the fact that the quantum spin is a true angular momentum \cite{edh}. Indeed, in the absence of a restoring force the system is predicted to behave as a gyroscope, exhibiting precession and nutation \cite{kimball1,kimball3}.

From now on, we assume that the system parameters are such that the cross-coupling terms are negligible. Assuming that the intrinsic rotation $\dot  \gamma$ is negligible, a sufficient condition is that $\omega_I \ll \omega_{\alpha,\beta}$. If the gyroscopic terms proportional to $\omega_I$ are significant but relatively small, the linear system of Eqs. (\ref{eqalpha}) and (\ref{eqbeta}) can still be diagonalized by a proper change of coordinates. On the other hand, when $\omega_I \gg \omega_{\alpha,\beta}$ the gyroscopic terms become dominant over the restoring terms, the linearized approximation breaks down, and one needs to analyze the full nonlinear gyroscopic dynamics as done elsewhere \cite{kimball1}. In Ref.~\cite{kimball4} an intermediate situation is analyzed in which the angle $\beta$ is librationally trapped and the $\alpha$ angle is free, leading to an interesting mixed librational-precessional regime.

Next, we evaluate the sensitivity of the levitated magnet to an external torque $\bm \tau$, in particular when this is induced by an external magnetic field $\bm B$. For negligible gyroscopic effects, the equations are simple decoupled harmonic oscillators. Without loss of generality we focus on the $\alpha$ degree of freedom and assume that an optimally oriented magnetic field $\bm B =B (0,1,0)$ is applied along the $y$-axis, as shown in Fig.~1a, leading to a torque $\bm \tau=\bm \mu \times \bm B$ along $z$. Moving to Fourier space, the equation of motion Eq.~(\ref{eqalpha}) becomes:
\begin{equation}
 \alpha(\omega)=\chi ( \omega )\tau_z( \omega ),
\end{equation}
where we have defined the librational susceptibility:
\begin{equation}
 \chi (\omega)=\frac{1}{I\left( -\omega^2+\omega_\alpha^2 + \frac{i \omega \omega_{\alpha}}{Q_\alpha} \right) },
\end{equation}
and 
\begin{equation}
\tau_z (\omega)=-\mu B (\omega)   \label{torquemuB}
\end{equation}
is the $z$ component of the torque.
By monitoring the librational motion we can infer the torque, and through Eq.~(\ref{torquemuB}) the applied field $B$. Thus we can measure $B$ through a purely mechanical measurement.

The only fundamental noise sources in a mechanical measurement are thermal mechanical noise and quantum detection noise \cite{braginsky}. Thermal noise, according to the fluctuation-dissipation theorem \cite{FD}, can be modeled by a one-sided spectral density:
\begin{equation}
 S_{\tau_T}=4 k_B T \textrm{Im} [ \chi(\omega)^{-1}]/\omega = 4 k_B T I \frac{\omega_\alpha}{Q_\alpha},
\end{equation}
where $k_B$ is the Boltzmann constant, $T$ the bath temperature and $\textrm{Im}$ stands for imaginary part. The dissipation represented by the imaginary part of $ \chi (\omega)$ includes all sources of mechanical dissipation, e.g., those arising from domain magnetic losses and magnons in the levitated magnet, which might set the ultimate limit for these types of sensors \cite{maglev}. 

For the torque quantum noise we adapt an approach usually applied to linear phase-insensitive force detection \cite{nimmrichter} (see Supplemental Information for a detailed derivation) resulting in the Standard Quantum Limit on torque:
\begin{equation}
S_{\tau_{SQL}} \geq 2 \hbar | \chi(\omega)|^{-1} = 2 \hbar I \left[  \left( -\omega^2 + \omega_\alpha^2 \right)^2 + \left( \frac {\omega \omega_\alpha}{Q_{\alpha}} \right)^2 \right]^{1/2}.  \label{SQL}
\end{equation}
The SQL is generally valid for any Heisenberg-limited linear motion detector, regardless of the implementation. Possible detection schemes could be SQUID-based \cite{maglev}, optical \cite{kimball2, timberlake, gieseler} or microwave \cite{merced}. In all cases, the SQL is in principle achievable. For SQUIDs, noise at factor less than 10 from the SQL has been demonstrated \cite{awschalom,squid}. It is worth noting that the SQL on mechanical measurements is not a fundamental limit either, similarly to the ERL, and in principle can be surpassed \cite{qnd}. However, it can be still considered a useful benchmark here, as we are interested in assessing if the SQL on mechanical detection implies the ERL in a ferromagnetic torque-based magnetometer. As shown in the following, we find that the SQL can be far below the ERL.

From Eq.~(\ref{torquemuB}) we can convert both thermal and SQL torque noise into an equivalent magnetic field noise. In terms of power spectral density the relation is simply $S_B=S_\tau/\mu^2$, where we are still assuming that the field is optimally oriented. In Fig.~2a we plot the SQL, the thermal noise, and the ERL noise Eq.~(\ref{ERL}), all expressed as magnetic field amplitude spectral density, $\sqrt{S_B}$, for realistic parameters demonstrated in Ref.~\cite{maglev}, i.e., a ferromagnetic NdFeB sphere with $R=30$ $\mu$m, levitated by the Meissner effect at $T=4.2$ K with a librational quality factor $Q=10^7$. For this setup we estimate the Einstein-de Haas frequency $f_I=\omega_I/2\pi=0.188$ Hz (vertical line in Fig.~2a). The librational frequency is set to $f_\alpha=\omega_\alpha/2\pi =10 f_I=1.88$ Hz. This value, one order of magnitude lower than in the current experiment, is expected to be achievable through a proper control over the trap tilt \cite{maglev}. We also show the thermal noise of a hypothetical advanced but yet realistic configuration with $Q=10^9$ and $T=50$ mK, and the spin projection noise for indipendent spins Eq. (1). It is apparent from Fig.~2a that with a mechanical detection at the SQL, the ERL is surpassed by many orders of magnitude, both on resonance and out of resonance. Moreover, the SQL detection noise is at least two orders of magnitude lower than the thermal noise, for experimentally demonstrated temperatures and quality factors. It can be seen that on resonance the SQL curve will also surpass the spin-projection noise for the same number of independent spins, confirming the results predicted in Ref. \cite{kimball1} for a gyroscope. This result is remarkable, as Eq. (\ref{SQLmag}) holds for independent particles, which is definitely not the case of a ferromagnet. Instead, for diluted spins with dipolar coupling, spin-projection noise is known to lead to the ERL \cite{mitchell2}.

\begin{figure}[!ht]
\includegraphics[width=8.6cm]{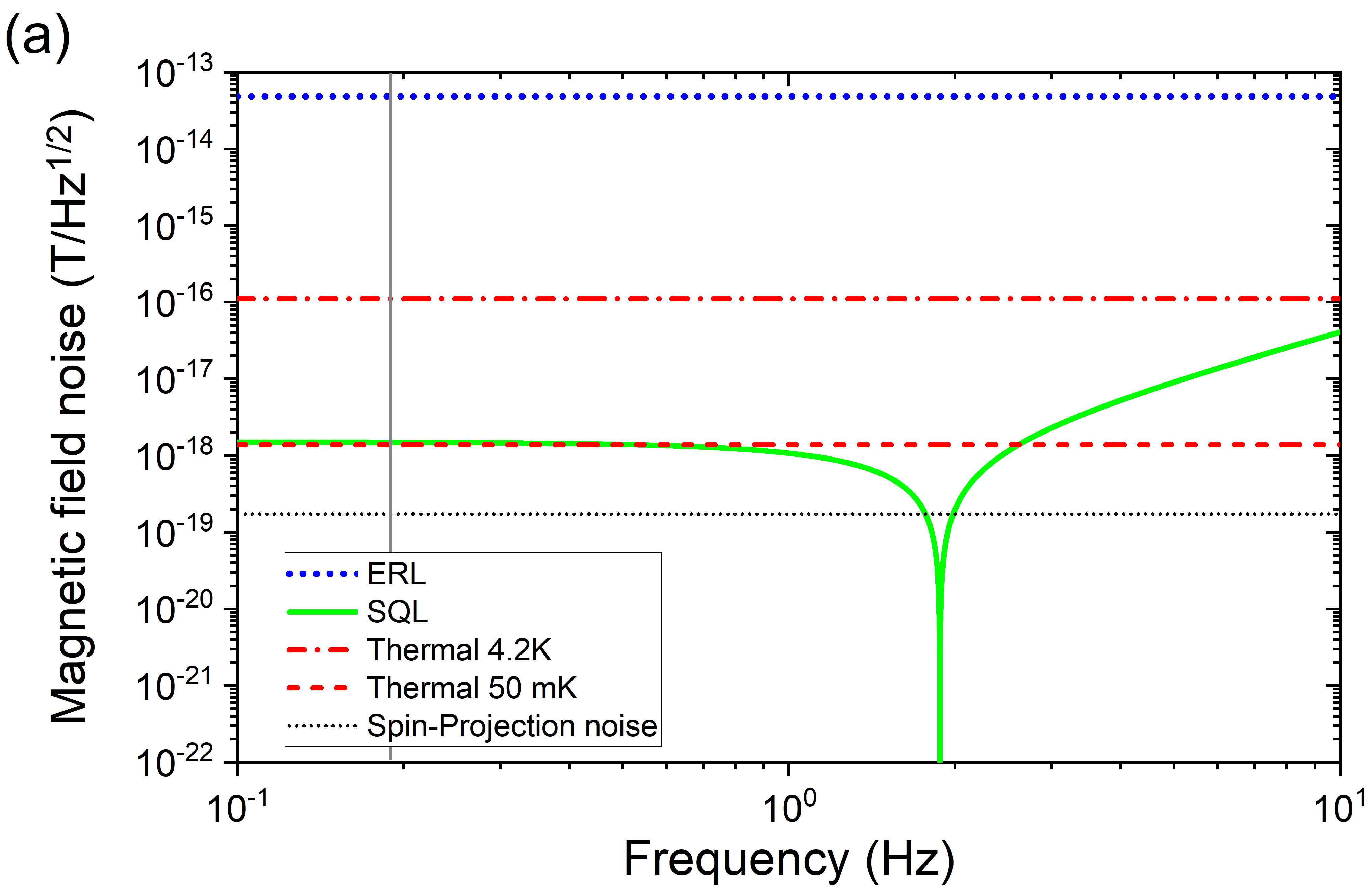}
 \includegraphics[width=8.6cm]{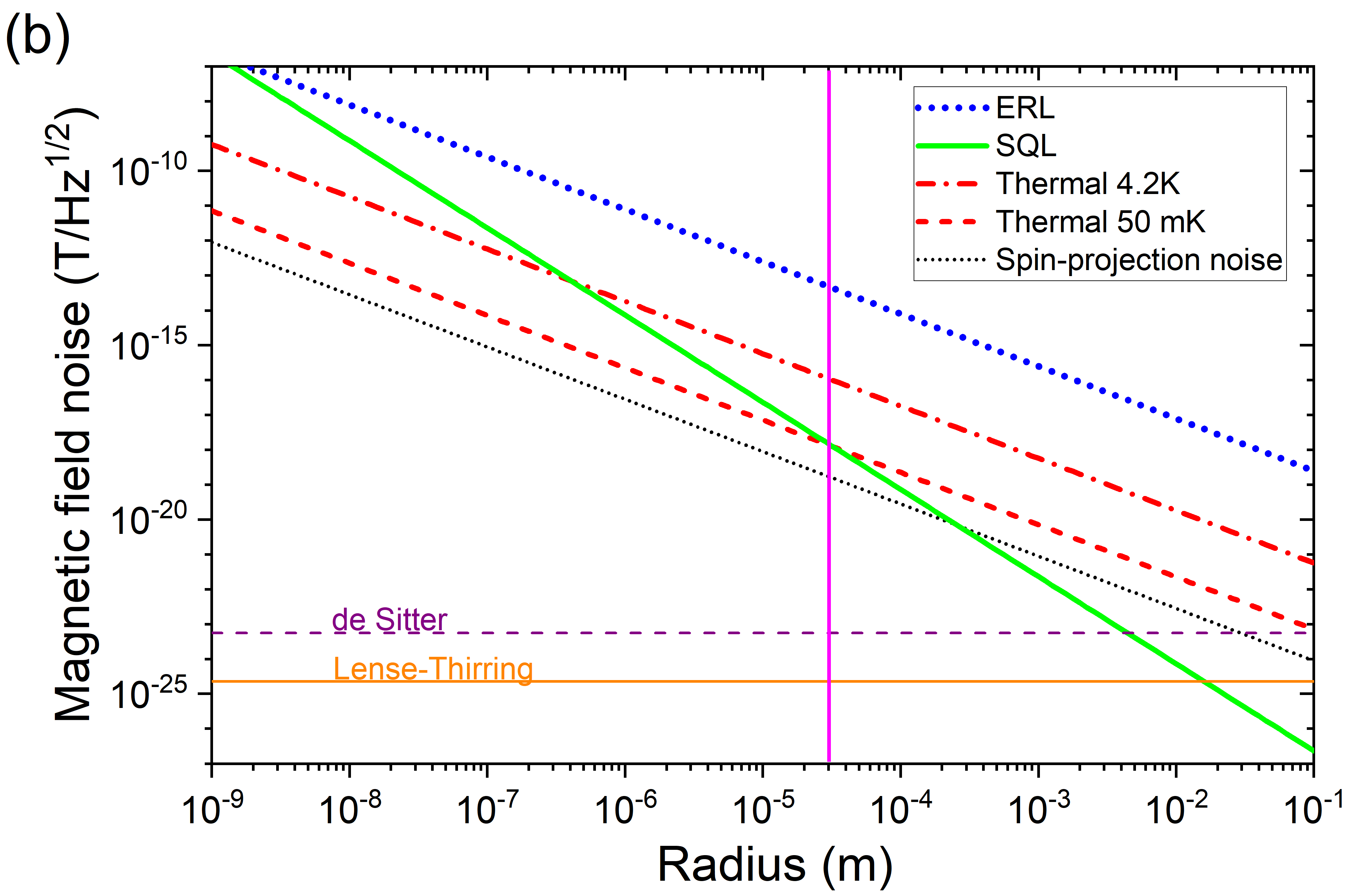}
 
\caption{(a) Magnetic field sensitivity as a function of frequency for a levitated NdFeB magnetic sphere with radius $R=30$ $\mu$m, $f_\alpha=1.88$ Hz: ERL (blue dotted line), thermal noise at  $Q=10^7$, $T=4.2$ K (red dashed-dotted) and at  $T=50$ mK, $Q= 10^9$ (red dashed), SQL on mechanical torque (green solid), spin projection noise for independent spins (thin black dotted). The vertical line represents the Einstein-de Haas frequency $f_I=0.1 f_{\alpha}$. (b) Subresonant magnetic field sensitivity as a function of radius $R$, with $f_\alpha=10 f_I$. Lines corresponding to ERL, thermal, SQL and spin-projection as in (a).  The vertical line marks the radius used in (a). Horizontal lines correspond to effective magnetic fields corresponding to relativistic frame dragging effects, as discussed in Ref.~\cite{kimball3}: de Sitter effect (purple dashed), Lense-Thirring effect (orange solid).} \label{plot}
\end{figure}

In Fig.~2b we illustrate the dependence of SQL, ERL and thermal noise on the radius $R$ of the levitated magnetic sphere. We set a constant ratio $f_\alpha/f_I=10$, the same value as in Fig. 2a. This choice implies $f_\alpha \propto R^{-2}$. 
The SQL is evaluated at subresonant frequency $f \ll f_\alpha$. 
Again, we can see from Fig. 2b that both SQL and thermal noise are well below the ERL. 

It is instructive to consider the scaling with the radius. Both thermal noise (assuming constant $Q$), ERL and spin-projection noise scale as $R^{-3/2}$, while the mechanical SQL scales as $R^{-5/2}$. Technically, the additional dependence of the SQL mechanical noise on $R$ is due to the different scaling with $R$ of the moment of inertia with respect to the volume (see Supplemental Information). Interestingly, the ERL and the SQL are comparable for $R\approx 1$ nm, i.e., at the scale where the ferromagnet approaches the single spin limit. This supports the heuristic explanation provided in Ref. \cite{kimball1}, where the ability to achieve sub-quantum-limited performance is attributed to the ferromagnet being composed of a large number $N$ of highly correlated spins, locked together on a well-defined direction by magnetic anisotropy. In this sense we expect the improvement over the ERL to increase steadily with $N$ and thus with increasing radius. 

Theoretically, the ratio between ERL and SQL increases indefinitely with increasing $R$. One can predict magnetic field resolution down to $10^{-24}$ T/$\sqrt{\textrm{Hz}}$ for $R=1$ cm, and $10^{-29}$ T/$\sqrt{\textrm{Hz}}$ for $R=1$ m. 
There is a practical limitation to the apparently unbounded improvement of resolution with size $R$: Fig. 2b is obtained by assuming that the librational frequency $f_\alpha$ can be locked to the Einstein-de Haas frequency $f_I$. The latter drops rapidly with increasing $R$, as $R^{-2}$. This means that ultrasensitive measurements with larger and larger $R$ become quickly unfeasible due to timescale reasons. For instance, for $R=1$ cm, the librational frequency becomes of the order 1 day$^{-1}$.
\begin{figure}[!ht]
\includegraphics[width=8.6cm]{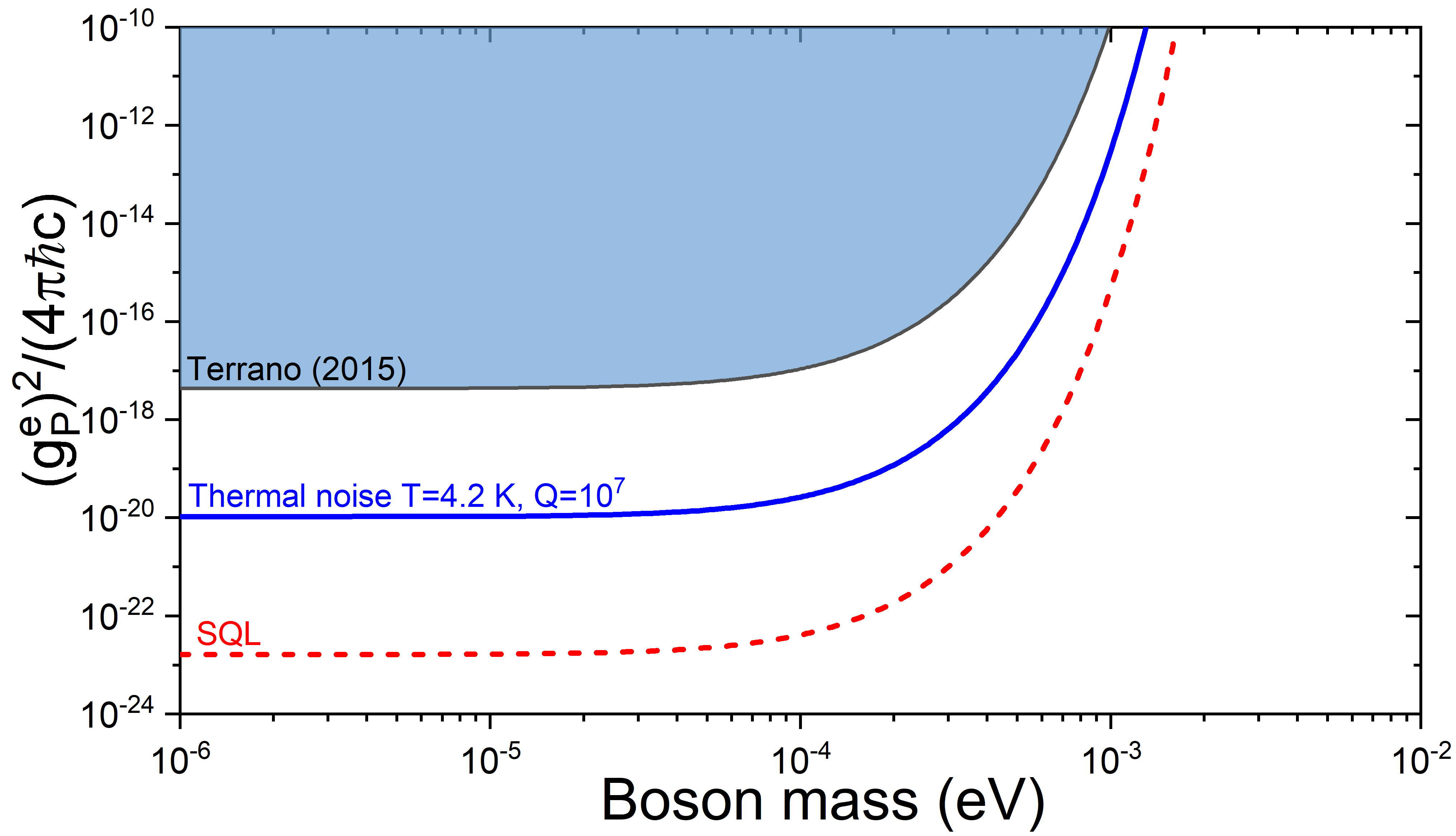} 
\caption{Exclusion plot from a hypothetical experiment searching for spin-spin interactions between electrons mediated by an exotic boson. The $x$ axis is the mass of the exchanged boson, the $y$ axis is the dimensionless spin-spin coupling \cite{kimball4}. We assume a levitated micromagnet with radius $R=0.2$ mm,   a rotating polarized sphere actuator of radius $2$ mm placed at distance $4$ mm underneath the levitated magnet, and measurement time $10^6$ s. Blue solid (red dashed) lines:  theoretical bound assuming thermal noise at $T=4.2$ K, $Q=10^7$ (SQL noise). Black line and shaded region: exclusion plot from the best experiment in the literature \cite{terrano}.}   \label{exclusion}
\end{figure}

Our findings indicate that a magnetomechanical system optimized for ultrasensitive torque measurements can outperform conventional magnetometers with the same volume by many orders of magnitude. For example, magnetometers based on spinor Bose-Einstein condensates (BEC) have reached the ERL for an effective dimension of $\approx 10$ $\mu$m \cite{stamperkurn}, and are expected to surpass the ERL \cite{mitchell} by employing spin squeezing techniques \cite{sewell} or using the ferromagnetic phase of the spinor condensate \cite{palacios}. However, these approaches do not come near the three to five orders-of-magnitude enhanced sensitivity of a levitated magnetomechanical system of similar size (see Fig. 2). This is because of the much higher density of correlated electron spins attainable in ferromagnets as compared to BECs (and other conventional magnetometers) and the rapid averaging of quantum and thermal noise via internal interactions \cite{kimball1}.

One promising field of application of a levitated torque sensor is the search for weak fundamental physics effects which are indirectly equivalent to an effective magnetic field, such as inertial frame rotations from general relativistic frame-dragging \cite{kimball3} or pseudomagnetic forces arising from exotic interactions beyond the standard model \cite{kimball4}. For the former case, we show for reference in Fig.~2b two horizontal lines corresponding respectively to the de Sitter effect (geodetic precession) and the Lense-Thirring effect (frame-dragging due to Earth rotation), as discussed in Ref.~\cite{kimball3}. Both effects can be modeled as a rotation of the local inertial reference frame with well-defined angular velocity $\Omega$, which is felt by the ferromagnet electron spins as an effective magnetic field $B_\textrm{eff}=\Omega/\gamma_0$. Figure~2a shows that a levitated ferromagnet in the tipping limit at SQL can achieve the sensitivity needed to resolve both effects on the timescale of 1 second for $R\approx 1$ cm. 

For the detection of pseudomagnetic forces, we consider the case of exotic spin-spin interactions between electrons arising from the exchange of an axionlike boson. In Fig.~\ref{exclusion} we illustrate the potential reach of an experiment based on a levitated magnet with $R=0.2$ mm, corresponding to a torsional frequency $f_\alpha=0.04$ Hz, and a rotating polarized source sphere with $2$ cm diameter placed at $4$ mm distance. We assume that the noise is limited either by thermal noise with $T=4.2$ K and $Q=10^7$, or by the SQL mechanical noise. Such an experiment would extend the current experimental bounds \cite{terrano} on the dimensionless coupling factor $g_p^2$ by three and six orders of magnitude in the two cases. 

In conclusion, we have shown that a ferromagnetic torque sensor based on a levitated magnet, operated at low enough thermal noise and readout close to the mechanical SQL, is able to overcome standard quantum limits on conventional magnetometers, and can surpass the ERL by many orders of magnitude. Remarkably, this goal appears within reach with existing technology, opening new prospects in fundamental physics experiments requiring extreme sensitivity.

\begin{acknowledgments}
We thank M.W. Mitchell and G. Gasbarri for discussions on the ERL. HU, CT and AV were supported by the EU H2020 FET project TEQ (grant 766900) and the Leverhulme Trust  (grant RPG-2016-046). DB was supported by the Cluster of Excellence {\it Precision Physics, Fundamental Interactions, and Structure of Matter} (PRISMA+ EXC 2118/1) funded by the German Research Foundation (DFG) within the German Excellence Strategy (Project ID 39083149), by the European Research Council (ERC) under the European Union Horizon 2020 research and innovation program (project Dark-OST, grant agreement No 695405), and by the DFG Reinhart Koselleck project. DFJK was supported by the US National Science Foundation under grant PHY-1707875. AOS was supported by US Department of Energy grant DESC0019450, the Simons Foundation grant 641332, and the US National Science Foundation grant PHY-1806557.
\end{acknowledgments}

\section*{Supplemental Material}

\setcounter{equation}{0}
\setcounter{figure}{0}
\renewcommand{\theequation}{S.\arabic{equation}}

\subsection*{Derivation of the Standard Quantum Limit on torque detection}

We consider an angle detector with imprecision noise on the angular measurement with (one-sided) spectral density $S_{\alpha}$ and a back-action torque noise with spectral density $S_{\tau_{BA}}$. For linear phase-insensitive detection the one-sided spectral densities satisfy the Heisenberg constraint $S_{\alpha}S_{\tau_{BA}} \geq \hbar^2$ . At a given frequency the imprecision noise can be converted into an equivalent torque noise $S_{\tau_\alpha} = S_{\alpha} |\chi (\omega)|^{-2}$. The Standard Quantum Limit (SQL) on torque is then achieved by minimizing the sum of the two contributions, which happens for $S_{\tau_{BA}}/S_{\alpha} = |\chi (\omega)|^{-2}$, yielding a total detection torque noise at the SQL given by Eq. (13) of main text.
In the subresonant limit $\omega \ll \omega_\alpha$, we can simplify this expression to $S_{\tau_{SQL}}\approx 2 \hbar I \omega_\alpha^2$.

\subsection*{Simplified heuristic argument to compare ERL-limited and SQL-limited performance, and scaling with $R$.}

A simple and intuitive derivation of the ERL can be found in Ref. \cite{mitchell}. Assume that a magnetic sensor directly measures the magnetic field 
$B_\mathrm{true}$ in a volume $V$, providing a measured value $B=B_\mathrm{true}\pm \delta B$ with uncertainty $\delta B$. Then the apparent magnetostatic energy in the volume $V$ can be formally written as:
\begin{equation}
\left\langle E \right\rangle  = \frac{{B_{\mathrm{true}}^2 V}}{{2\mu _0 }} + \frac{{\left\langle {\delta B^2 } \right\rangle V}}{{2\mu _0 }},
\end{equation}
where the second term expresses the energy resolution associated to the measurement uncertainty. Allowing for time averaging with integration time $t$, one can define a quantity $\left\langle {\delta B^2 } \right\rangle Vt/\left( 2\mu _0 \right) $ with units of action, where $\left\langle {\delta B^2 } \right\rangle t =  S_B$ can be interpreted as a magnetic field power spectral density. By formally equating the time-averaged energy resolution to $\hbar$ one obtains the ERL.

However, in our scheme we perform a mechanical rather than a magnetic measurement. In order to apply the same argument we need to define the apparent mechanical energy associated with the measurement. Application of a magnetic field $B$ produces a torque $\tau=\mu B$ and an angular deflection at subresonant frequency $\alpha=\tau/\kappa$ with $\kappa=I \omega_\alpha^2$. The associated mechanical energy is:
\begin{equation}
E = \frac{1}{2}\kappa \alpha ^2  = \frac{{\tau ^2 }}{{2\kappa }} = \frac{{\mu ^2 }}{{2\kappa }}B^2.
\end{equation}

Following the same argument as for the ERL, we can write the apparent energy associated with the measurement as:
\begin{equation}
\left\langle E \right\rangle  = \frac{{\mu ^2 }}{{2\kappa }}B_{true}^2  + \frac{{\mu ^2 }}{{2\kappa }}\left\langle {\delta B^2 } \right\rangle .
\end{equation}
The quantum limit can now be derived by relating the energy resolution per unit bandwidth to $\hbar$, as:
\begin{equation}
\frac{{\mu ^2 }}{{2\kappa }}\left\langle {\delta B^2 } \right\rangle t \ge \hbar ,
\end{equation}
which leads to:
\begin{equation}
S_B  \ge \frac{{2\hbar I\omega _\alpha ^2 }}{{\mu ^2 }}.
\end{equation}
This is the same expression of SQL noise that one obtains from Eq.~({\ref{SQL}}) in the subresonant limit.

Note that, since we have assumed that $\omega_\alpha \propto f_I \propto V/I$, and taking into account that the magnetic moment $\mu$ scales with the volume $V$, the SQL noise is found to scale as $1/I \propto R^{-5}$. This explains the different scaling of SQL noise with $R$ with respect to the ERL and the ideal spin projection noise for $N$ independent particles, both scaling with $R^{-3}$.


\begin{thebibliography}{<99>}

\bibitem{kimball1} D.F. Jackson Kimball, A.O. Sushkov, and Dmitry Budker, Precessing Ferromagnetic Needle Magnetometer, Phys. Rev. Lett. 116, 190801 (2016).

\bibitem{mitchell} M.W. Mitchell, S. Palacios Alvarez, Quantum limits to the energy resolution of magnetic field sensors, Rev. Mod. Phys. 92, 021001 (2020).

\bibitem{MFM} Y. Martin, and K. Wickramasinghe, Magnetic Imaging by Force Microscopy with 1000 $\AA$ Resolution, Appl. Phys. Lett. 50, 1455 (1987).

\bibitem{MRFM} O. Zuger, and D. Rugar (1993), First images from a magnetic resonance force microscope, Appl. Phys. Lett. 63, 2496 (1993).

\bibitem{kimball3} P. Fadeev, T. Wang, Y.B. Band, D. Budker, P.W. Graham, A.O. Sushkov, and D.F. Jackson Kimball, Gravity Probe Spin: Prospects for measuring general-relativistic precession of intrinsic spin using a ferromagnetic gyroscope, Phys. Rev. D 103, 044056 (2021).

\bibitem{kimball4} P. Fadeev, C. Timberlake, T. Wang, A. Vinante, Y.B. Band, D. Budker,  A.O. Sushkov, H. Ulbricht, D.F. Jackson Kimball, Ferromagnetic Gyroscopes for Tests of Fundamental Physics, Quantum Sci. Technol. 6, 024006 (2021). 

\bibitem{braginsky} V.B. Braginsky, Classical and quantum restrictions on the detection of weak disturbances of a macroscopic oscillator, Sov. Phys. JETP 26, 831 (1968). 

\bibitem{kimball2} T. Wang, S. Lourette, S.R. O Kelley, M. Kayci, Y.B. Band, D.F. Jackson Kimball, A.O. Sushkov, and D. Budker, Dynamics of a ferromagnetic particle levitated over a superconductor, Phys. Rev. Appl. 11, 044041 (2019).

\bibitem{gieseler} J. Gieseler, A. Kabcenell, E. Rosenfeld, J.D. Schaefer, A. Safira, M.J.A. Schuetz, C. Gonzalez-Ballestero, C.C. Rusconi, O. Romero-Isart, M.D. Lukin, Single-Spin Magnetomechanics with Levitated Micromagnets, Phys. Rev. Lett. 124, 163604 (2020).

\bibitem{maglev} A. Vinante, P. Falferi, G. Gasbarri, A. Setter, C. Timberlake and H. Ulbricht, Ultralow Mechanical Damping with Meissner-Levitated Ferromagnetic Microparticles, Phys. Rev. Applied 13, 064027 (2020).

\bibitem{romero} C.C. Rusconi, V Pochhacker, K Kustura, J.I. Cirac, O. Romero-Isart, Quantum spin stabilized magnetic levitation, Phys. Rev. Lett. 119, 167202 (2017).

\bibitem{edh} A. Einstein and W.J. de Haas, Experimenteller Nachweis der Ampereschen Molekularstrome, Dtsch. Phys. Ges. Verh. 17, 152 (1915).

\bibitem{FD} H.B. Callen and T.A. Welton, Irreversibility and generalized noise, Phys. Rev. 83, 34 (1951).

\bibitem{nimmrichter} S. Nimmrichter, K. Hornberger, and K. Hammerer, Optomechanical Sensing of Spontaneous Wave-Function Collapse, Phys. Rev. Lett. 113, 020405 (2014).

\bibitem{timberlake} C. Timberlake, G. Gasbarri, A. Vinante, A. Setter, H. Ulbricht, Acceleration sensing with magnetically levitated oscillators, Appl. Phys. Lett. 115, 224101 (2019).

\bibitem{merced} N.K. Raut, J. Miller, J. Pate, R. Chiao, J.E. Sharping, Meissner levitation of a millimeter size neodymium magnet within a superconducting radio frequency cavity, IEEE Trans. Appl. Supercon. 31, 1500204 (2021).

\bibitem{awschalom} D.D. Awschalom, J.R. Rozen, M.B. Ketchen, W.J. Gallagher,
A.W. Kleinsasser, R.L. Sandstrom, and B. Bumble, Low noise modular microsusceptometer using nearly quantum limited dc SQUIDs, Appl. Phys. Lett. 53, 2108 (1988).

\bibitem{squid} P. Falferi, M. Bonaldi, M. Cerdonio, R. Mezzena, G.A. Prodi, A. Vinante, S. Vitale, 10 $\hbar$ superconducting quantum interference device amplifier for acoustic gravitational wave detectors, Appl. Phys. Lett. 93, 172506 (2008).

\bibitem{qnd} V.B. Braginsky and Y.I. Vorontsov, Quantum-mechanical limitations in macroscopic experiments and modern experimental technique, Sov. Phys. Usp. 17, 644 (1975).

\bibitem{mitchell2} M.W. Mitchell, Scale-invariant spin dynamics and the quantum limits of field sensing, New J. Phys. 22 053041 (2020).

\bibitem{stamperkurn} M. Vengalattore, J.M. Higbie, S. R. Leslie, J. Guzman, L.E. Sadler, and D.M. Stamper-Kurn, High-Resolution Magnetometry with a Spinor Bose-Einstein Condensate, Phys. Rev. Lett. 98, 200801 (2007).

\bibitem{sewell} R.J. Sewell, M. Koschorreck, M. Napolitano, B. Dubost,  N. Behbood, and M.W. Mitchell, Magnetic sensitivity beyond the projection noise limit by spin squeezing, Phys. Rev. Lett. 109, 253605 (2012).

\bibitem{palacios} S. Palacios, S. Coop, P. Gomez, T. Vanderbruggen, Y.N.M. de Escobar, M. Jasperse, and M.W. Mitchell, Multi-second magnetic coherence in a single domain spinor Bose–Einstein condensate, New J. Phys. 20, 053008 (2018).

\bibitem{terrano}  W.A. Terrano, E.G. Adelberger, J.G. Lee, B.R. Heckel, Short-Range Spin-Dependent Interactions of Electrons: A Probe for
Exotic Pseudo-Goldstone Bosons, Phys. Rev. Lett. 115, 201801 (2015).

\end{thebibliography}
\end{document}